\documentclass{epl}

\title{Ab initio study of surface stress response to charging}
\author{Yoshitaka Umeno\inst{1,2} \and Christian Els\"asser\inst{3}
  \and Bernd Meyer\inst{4} \and Peter Gumbsch\inst{1,3} \and
  Martina Nothacker\inst{5} \and J\"org Wei{\ss}m\"uller\inst{5,6}
  \and Ferdinand Evers\inst{5,7}}

\institute{
 \inst{1}Institut f\"ur Zuverl\"assigkeit von Bauteilen und Systemen, University of Karlsruhe, Karlsruhe, Germany\\
 \inst{2}Institute of Industrial Science, The University of Tokyo, Tokyo, Japan\\
  \inst{3}Fraunhofer-Institut f\"ur Werkstoffmechanik IWM, Freiburg, Germany\\
  \inst{4}Institut f\"ur Theoretische Chemie, Ruhr-Universit\"at Bochum, Bochum, Germany \\
  \inst{5}Institut f\"ur Nanotechnologie, Forschungszentrum Karlsruhe, Karlsruhe, Germany \\
  \inst{6}Technische Physik, Universit\"at des Saarlandes, Saarbr\"ucken,
Germany\\
  \inst{7}Institut f\"ur Theorie der Kondensierten Materie, Universit\"at Karlsruhe, Karlsruhe, Germany
}

\begin{document}

\maketitle

\begin{abstract}
We explore an efficient way to numerically evaluate the response of
the surface stress of a metal to changes in its superficial charge
density by analysis of the strain-dependence of the work function of
the uncharged surface. As an application we consider Au(111), (110)
and (100) surfaces, employing density functional calculations. The
sign of the calculated response parameter can be rationalized with the
dependence of the surface dipole and the Fermi energy on strain. The
numerical value falls within the range indicated by experiment. The
magnitude can explain the experimentally observed
volume changes of nanoporous materials upon charging.
\end{abstract}

Recent experiments reveal a macroscopic expansion or contraction when
high surface area metals with nanometer-sized porosity are
electrically charged, the signature of changes in the surface bond
forces when a space-charge layer is generated
\cite{Joerg_Science,NPG_Kramer}. The parameter which quantifies these
forces in the continuum description of solid surfaces, the surface
stress, $f$, is a topic of current interest in surface science, since
it is intricately related to the surface electronic structure and
bonding, and since it is relevant for reconstruction as well as for
the stress in thin film devices
\cite{Surface_Ibach,SurfaceCammarata}. Furthermore, the impact of the
energy-conjugate quantity to $f$, the tangential strain $e$, on the
surface electronic structure is being recognized as central for the
catalytic activity of metal surfaces \cite{Norskov1997,Kolb2005}. The
response of $f$ to changes in the superficial excess charge density
(per area), $q$, relates to fundamental issues in electrochemistry,
such as electron transfer in surface-adsorbate bonds or microscopic
processes in the electrochemical double layer
\cite{Surface_Ibach,Surface_Haiss1, Lipkowski,Surface_Friesen}.
Experiments finding a stronger response when there is less adsorption
point towards the effect of $q$ on the bonding in the metal surface as
a decisive factor \cite{Surface_Haiss2,NPP_Viswanath}. Yet, the
microscopic processes linking $q$ to forces and relaxation at metal
surfaces are poorly understood. A recent density functional theory
(DFT) simulation of Au cluster ions highlights the role of stretch
(relaxation of the surface in normal direction) in response to charge
\cite{WeigendCluster}. Qualitatively, the relaxation might be
rationalized as driven by out-of-plane Hellman-Feynman forces between
the excess charge and the surface layer of atoms. It was argued that
the outward relaxation feeds back into the (in-plane) surface stress
owing to the transverse contraction tendency of solids
\cite{WeigendCluster}. This hypothesis, however, awaits verification
by experiments or
simulation using single crystal surfaces of different orientation.

Here, we go a first step into this direction and explore the response
of $f$ to $q$ for surfaces of gold. We begin by outlining an {\it ab
  initio} approach, which allows to investigate the response with good
accuracy. In close analogy to a procedure which has recently been
proposed for experimental work \cite{Surface_Vasiljevic}, we exploit a
Maxwell relation that permits studying the response without explicit
introduction of an excess charge into the calculation. For a general
discussion of Maxwell relations with an eye on applications to
interfaces, we refer the reader to the textbook by Ibach,
Ref. \cite{Ibach_Textbook_Surfaces_2006}. Here, we apply the concept to gold surfaces
using DFT calculations, and we give a brief discussion of the
relevant microscopic processes.

As a model situation, we consider a metal plate extended laterally and
of finite thickness. The energetics of the extended bulk metal and
the excess at the surface are represented by energy densities,
$\Psi$ and $\psi$, respectively. In experimental studies and in
continuum theory the densities are commonly measured per volume or
per area, whereas {\it ab initio} studies often specify
energies per atom or per crystallographic unit cell.
Both conventions may be treated by a common set of equations, provided
that the surface area, $A$, and plate thickness, $d$, of the continuum picture are
measured in coordinates of the {\it undeformed} lattice (Lagrange coordinates;
see Ref. \cite{Balance_Weissmueller} and references therein), a
convention which we adopt here.
More specifically, we take $\Psi$ and $\psi$ to obey
\begin{equation}
  \label{e1}
     E_d(e,q) = A \, d \, \Psi(e) + A \,  \psi_d(e,q)
\end{equation}
where $E$ is the plate's total energy and the subscript - which we
shall drop henceforth for brevity - accounts for a possible dependency
of effective surface properties on the number of layers in thin
plates. We allow for the plate to be elastically strained in the
tangent plane, while it is free to relax along its
normal, with no external forces acting on its surface. The tangential
strain, $e$, represents the relative change in surface area - as measured in
laboratory coordinates - owing to the in-plane elastic deformation. Our
particular interest is in the dependence of $\psi$ on $q$ and on
$e$. A scalar surface stress is defined via
\begin{equation}
  \label{e2}
f(e,q)= \partial\psi / \partial e |_{q}.
\end{equation}
Generally, the forces by which the surface interacts with the
underlying bulk crystal are anisotropic. For planar surfaces they can
be described by a $2\times 2$ tensor, of which $f$ is half the trace. For
reference below we note that, by symmetry, the surface stress is isotropic on
the (111) and (100) surface of an fcc lattice, but it may be anisotropic on (110.
\footnote{
To include the anisotropy, let $\Psi$ and $\psi$ depend on the
tangential strain tensor, $\bf{e}$, instead of $e$
\cite{Balance_Weissmueller}. The surface stress is here obtained as a
second rank tensor in the plane, $ {\bf s}({\bf e},q)= \partial \psi /
\partial {\bf e} |_{q}$. The stress-charge coefficient of Eq. (\ref{e
  varsigma}) is then also a second rank tensor,  $\mbox {\boldmath $\varsigma$}= \left. \partial {\bf s} / \partial q \right|_{ \bf e} $
. The appropriate form of the Maxwell relation, Eq. (\ref{e5}), is
$\partial {\bf s} /
\partial q |_{\bf e} = -{q_0}^{-1} \partial\mu / \partial {\bf e}|_{q}$. }

The response of $f$ to charging defines a scalar surface-stress-charge
coefficient \cite{Balance_Weissmueller},
\begin{equation}
    \label{e varsigma}
    \varsigma = \left. \frac{\partial f}{\partial q}\right|_{e} =
\frac{\partial}{\partial q}\left[ \left.\frac{\partial \psi}{\partial e}\right|_q\right]_e.
\end{equation}
Let us introduce the electron chemical potential, $\mu=\partial E /
\partial n|_e$ with $n = - A q / q_0$ the number of electrons in the
system, where $q_0$ denotes the elementary charge, $q_0{=}1.6022
\times 10^{-19}$ C. By inspection of Eq. (\ref{e1}) it is
seen that
\begin{equation}
    \mu =-q_0  \partial\psi / \partial q|_e .
    \label{e4}
\end{equation}
The Maxwell relation for $\psi(e,q)$,
\begin{equation}
  \label{e5}
    \partial f / \partial q |_{e} = -{q_0}^{-1} \partial\mu / \partial e|_{q},
\end{equation}
relates $\varsigma$ to a change in $\mu$ when the surface is strained tangentially at constant $q$.

The work function, $W$, is related to $\mu$ by
\begin{equation}
    \mu = -W = \phi_{\infty} - \varepsilon_F,
    \label{eq:wf}
\end{equation}
where $\phi_{\infty}$ is the asymptotic vacuum value of the
electrostatic potential and $\varepsilon_F$ is the energy of the
highest occupied single particle state (Fermi energy)
of the metal plate \cite{Smoluchowski_work_function}.
Therefore, as an immediate consequence of
Eq. (\ref{e5}), the linear response quantity $\partial f/\partial
q|_{q=0}$ can be obtained by calculating the work function change when
the uncharged slab is strained uniformly in the plane, and using Eqs. (\ref{e5}) and (\ref{eq:wf}). This is a routine calculation with
standard codes. The advantages are twofold: a)
Extending codes to include excess surface charges is
not required at this point.
b) Directly determining $f$ using the definition
Eq. (\ref{e2}) comes with technical difficulties: in order to
obtain $\psi$ from total energy computations, total energies from
slab-calculations with different thicknesses are to be compared.
Taking finite energy differences between cells with
different sizes keeping very high numerical accuracy
(needed for derivatives) can be computationally expensive,
because the number of basis functions varies with the
cell size. When using Eq. (\ref{e5}),
explicit knowledge of the free energy is not required.

\begin{table}
\caption{\label{t1}
    Properties of unreconstructed
    Au-surfaces at $e = 0$ as obtained in this work: free energy, $\psi$, surface stress,
    $f$, work function, $W$, and surface stress response, $\varsigma$. $\psi$ and $f$
    have been determined via Eqs. (\ref{e1}) and (\ref{e2}),
     extrapolating $1/d\to 0$. Published values obtained using LDA are shown for comparison,
     and are marked by reference.
   }
\begin{center}
\begin{tabular}
{c|c|c|c|c}
\hline\hline
      & $\psi$ [J/${\rm m}^2$] ([eV/cell]) &  $f$ [J/${\rm m}^2$] & $W$ [eV]  & $\varsigma$ [V]
      \\\hline
(111) &  1.114 (0.502)  &   3.317     &  5.65     &  $-1.86$              \\
      &  $1.25^{\mbox{\cite{Needs}}}$ $1.39^{\mbox{\cite{Gala,Gala2}}}$ &
      $2.77^{\mbox{\cite{Needs}}}$  & $5.63^{\mbox{\cite{Fall}}}$& \\
      &  $1.68^{\mbox{\cite{Kollar}}}$  &   $2.56^{\mbox{\cite{Kollar}}}$     &           &     \\
(100) &  1.343 (0.700)  &   2.723     &  5.56     &  $-0.90$               \\
      &  $1.62^{\mbox{\cite{Gala,Gala2}}}$ $1.44^{\mbox{\cite{Fio}}}$ &
      $3.11^{\mbox{\cite{Fio}}}$    &  $5.61^{\mbox{\cite{Fio}}}, 5.53^{\mbox{\cite{Fall}}}$  &
      \\
(110) &  1.372 (1.010)  &   2.020     &  5.42     &  $\approx 0$                           \\
      &  $1.75^{\mbox{\cite{Gala,Gala2}}}$           &         &   $5.41^{\mbox{\cite{Fall}}}$    &
  \\\hline\hline
\end{tabular}
\end{center}
\end{table}

\begin{figure}[t]
    \onefigure[width=13cm]{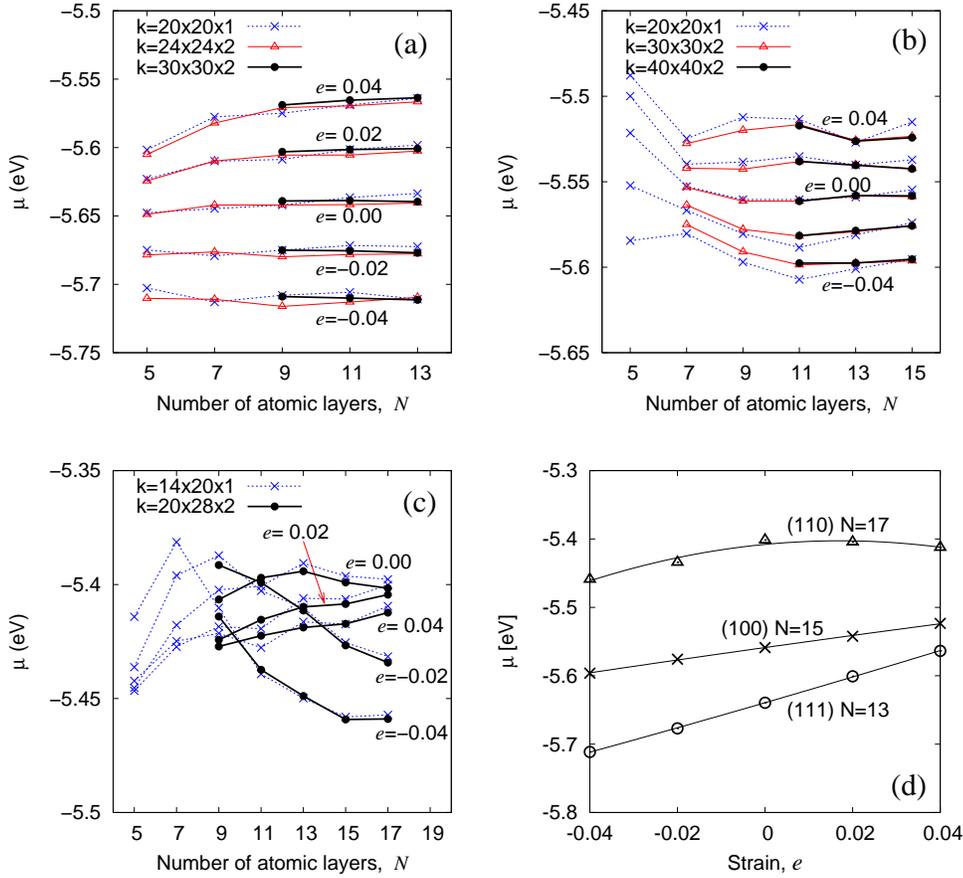}
    \caption{(color online) (a)-(c) Electron chemical potential,
    $\mu$,of the Au(111), (100) and (110) slab models at various
    tangential strains, $e$, as a function of the number of atomic
    layers  $N$. (d) $\mu$ as a function of $e$ for the thickest
    samples in each orientation. Lines: fit by linear (111, 100) and
    quadratic (110) functions.}
    \label{f1}
\end{figure}

Next, we apply our method and determine $\varsigma$ for Au
surfaces (111), (100) and (110) from Eq. (\ref{e5}). To this end we
performed DFT calculations within LDA \cite{Hedin-Lundqvist}
using a mixed-basis pseudopotential (MBPP)
approach \cite{MBPP1,MBPP2,MBPP6,MBpp7}.
The core-valence interactions in Au were described by norm-conserving ionic pseudopotentials,
and the valence
states were represented by a mixed basis of plane waves ($E_{pw}$ = 20 Rydberg) and five local
d-orbitals per Au atom ($r_{lo}$ = 2.5 Bohr).
Monkhorst-Pack k-point meshes were used for Brillouin-zone
integrations \cite{Monkhorst-Pack}. Our present computational LDA-MBPP
approach is the same as for earlier ab-initio studies of gold surfaces
by Ref. \cite{Bohnen1998} (and references therein).
We consider a slab model with periodic
boundary  conditions in all dimensions, containing
two surfaces exposed to vacuum with an odd number, $N$,
of atomic layers (5, 7, 9, $\cdots$)
and a vacuum region thicker than 13 \AA, which corresponds to at least
6 atomic layer thicknesses. (We have concentrated on the case $N$ odd,
to avoid even-odd effects, which impair the extrapolation $N\to\infty$.)
The thick vacuum region is introduced to prevent undesirable interactions of the surfaces
via the periodic boundary conditions.
As a reference configuration for the strain we
take the film with lattice parameter, $a$, identical to the computed
equilibrium value of bulk Au, $a_0$ = 4.085\AA. The isotropic in-plane strain $e$ is
applied (i.e., equal strain components of magnitude $e/2$ along two
orthogonal directions in the plane, giving a relative area change of
magnitude $e$), and the layers are relaxed, keeping their in-plane
dimensions fixed while allowing free relaxation out-of-plane,
until all the relevant atomic forces are below 0.01 eV/\AA.

The DFT calculations directly give the value $\varepsilon_F$ of the
highest occupied electronic level.
The electrostatic potential in the vacuum region saturates
rapidly with increasing
distance to the metal surface. Thus, the value for the electrostatic
potential at the center of
the vacuum layer gives an excellent estimate for $\phi_{\infty}$.
Then, $W$ and $\mu$ follow from Eq. (\ref{eq:wf}). Table 1 compares published LDA-data for $\psi$, $W$, and $f$ for
surfaces of different orientation to our values.
It is seen, that the trends agree well.

Figure \ref{f1}(a) shows $\mu$ as a function of $N$ for (111)-oriented slabs for various $e$.
The most obvious observation is that $\mu$ varies systematically, and essentially linearly, with $e$.
Fig. \ref{f1}(a) also reveals an interesting detail:
the work function ($\mu=-W$) at nonzero strain
is very sensitive to the layer thickness in thin slabs. In particular when $N$ is doubled from
$5$ to $11$ at $e=0.04$, the resulting shift can be as important, as the
change induced by doubling the strain.
The data converge reasonably well for $N{\geq}9$ atomic layers (20 \AA).
For the k-point mesh density, convergence is
achieved with 24$\times$24$\times$2 mesh points.

The values of $\mu$ for the thickest film, $N {=} 13$,
are plotted over the strain, $e$, in Fig. \ref{f1}(d).
The linear variation is confirmed, and the slope is determined as $\varsigma = -1.86\pm0.016$V.
The sign is negative, indicating that the slab tends to expand laterally when
positively charged. (Recall that a positive $f$
represents a tendency of the surface to laterally
compress the underlying crystal.)

Our evaluation of the Au(100) surface follows exactly the same lines
and $\mu$ has been plotted in Fig. \ref{f1}(b). Again, there is a linear behavior, and we obtain $\varsigma =-0.90\pm0.02$V, a factor of two
smaller than for Au(111).

In contrast to the previous cases the electron
chemical potential of the (110) surface
is very difficult to obtain. The thick layer limit is not reached,
even at $N{=}17$, as shown in Fig. \ref{f1}(c).
The strain-dependence of $\mu$ for the (110) films
differs noticeably from that of the (111) and (100) films. Taken at face value,
the data (Fig. \ref{f1}(d)) imply a quadratic,
rather than a linear variation.

The most important systematical source of error in our calculation is due to
the approximation in the exchange correlation functional used for DFT.
We have used LDA here, which is a
controlled approximation for the homogenous electron gas, i.e. in the
absence of density inhomogeneities.
It is well known, that the application of LDA to vacuum surfaces is
not controlled, in the sense that the expansion parameter
(density gradient times Fermi wavelength) is no longer small. Therefore
it is not clear that gradient terms leave unaffected the
quantitative results presented in Table \ref{t1}.
Further calculations testing
functional dependencies are necessary and our future work is devoted
to this issue.

Despite of this disclaimer,
we believe that, qualitatively, our findings are generally valid
and representative of (unreconstructed) generic gold surfaces. The Maxwell relation, Eq. (\ref{e5}), is exact, and applies irrespective of the layer thickness.
It rigorously shows that the sign of $\varsigma$
is negative, as long as the electron chemical
potential grows with an {\it in}creasing strain in the plane, $e$.
This variation may be the superposition of several contributions
a more detailed analysis of which has to be relegated to a forthcoming publication \cite{YUJWFEunpublished}.
We mention here the most obvious terms, which cooperate to shift $\mu$ in the same direction, implying $\partial \mu/\partial e{>}0$.

In order to understand the first mechanism, recall that the surface dipole describes how the electronic charge relaxes at a crystal surface after cleavage. This relaxation increases
the work function $W$ \cite{Smoluchowski_work_function}. In the laterally compressed ($e<0$) material, the increase in electron density and the normal expansion of the surface dipole (due to the transverse elastic response of the lattice) cooperate to increase the dipole strength, thereby increasing $W$ and reducing $\mu$.

Secondly, the trend $\partial \mu/\partial e{>}0$ may also be understood in the framework of the microscopic picture proposed in Ref. \cite{WeigendCluster}, which considers the relaxation of the ion cores. Just as excess electronic charge ($\delta q < 0$) on the surface of Au acts to displace the ion cores outwards relative to their Wigner-Seitz cell, resulting in an tendency for in-plane contraction ($\delta f > 0$), the converse effect is expected when the surface is compressed laterally ($\delta e < 0$): the ion cores will relax outward, closer to the surface of the electronic charge distribution. This trend may equivalently be viewed as the formation of an ionic dipole, on top of the electronic one. It again acts to increase the net dipole strength ($\delta \mu <0$).

Thirdly, we note that the strain has also important consequences for
the band structure, where it induces a band shift and a band
flattening. For instance, the strain induced shift of the $d$-band
center of gold relative to $\varepsilon_F$ is large. It can exceed the
shift of $\mu$ found in our work by one or two orders of magnitude\cite{Norskov1997}. However, of interest to us is the change in $\varepsilon_F$ with respect to the vacuum level. For this quantity, effects of band shift and flattening nearly compensate each other.

The above effects depend on the atomic structure of the surface and, hence, on the orientation. This is consistent with the considerable difference in the values of $\varsigma$ for (111) and (100). The observed trend, larger response on (111), agrees qualitatively with that of the surface {\it stretch} response to surface charge reported in Ref. \cite{WeigendCluster}, consistent with the suggested coupling between surface stress and stretch.

The (110) surface presents a special case. In principle, it is possible that its parabolic $\mu (e)$ carries information about real metallic films. However, we feel that our results are not conclusive here because of the issues of convergence. It is conceivable that the slow convergence with film thickness signals the instability of the unreconstructed surface studied in our work with respect to a (missing row) reconstruction. An alternative reason could relate to the low symmetry of the (110) surface: the surface stress and its response to charging may well be anisotropic. By definition, the scalar $\varsigma$ is the mean of the two diagonal elements $\mbox {\boldmath $\varsigma$}_{xx}$ and $\mbox {\boldmath $\varsigma$}_{yy}$ of a tensor response parameter. The small value of $\varsigma$ which might be extracted from the data in Fig. \ref{f1}(d) could therefore be due to mutual cancelation of larger contributions $\mbox {\boldmath $\varsigma$}_{xx}$,$\mbox {\boldmath $\varsigma$}_{yy}$ of opposite sign.

Electrochemical experiments on Au (111) suggest values of $\varsigma$ in the range -0.67 to -0.9 V \cite{Surface_Haiss2,Ibach Au111Au100}. The most negative values are observed for more weakly adsorbing electrolytes, where a smaller fraction of the total charge $q$ is transferred away from the metal surface and into bonds with adsorbed ions \cite{Surface_Haiss2}. In fact, experiments on Pt surfaces find $\varsigma$ to decrease from -1.0 V to apparent saturation at -1.8 V when the adsorption is progressively reduced in dilution series \cite{NPP_Viswanath}. The ``intrinsic'' magnitude of $\varsigma$ -- in the absence of adsorption -- may therefore be larger by as much as a factor of two as compared to experimental values recorded at typical concentrations. In this respect it is also remarkable that experimental data for the variation of the potential of zero charge of Au(111) as a function of lateral strain -- analogous to the present numerical procedure of computing $\partial\mu/\partial e$ -- have been invoked to estimate the value of $-q_0^{-1} \partial\mu/\partial e$ at -2 V \cite{Surface_Friesen}, similar to our numerical result. We therefore conclude that our result for Au(111) is consistent with the -- limited -- experimental data available at present. In view of the lack of fundamental understanding it is noteworthy that the computation for Au in vacuum reproduces the experimental sign of $\varsigma$ of Au in electrolytes, and that our results for $\varsigma$ are of the correct order of magnitude to explain the experimentally observed volume changes of nanoporous materials upon charging \cite{NPG_Kramer,Joerg_Science}. This supports the - not undisputed \cite{Surface_Haiss1,NPP_Viswanath,Surface_Friesen} - notion of a large intrinsic value of $\varsigma$, related to processes and forces within the metal surface rather than to adsorption.

In conclusion, an efficient way to compute the surface stress response to surface charging by investigation of the electron chemical potential of strained but uncharged surfaces is presented and its application to Au surfaces by means of {\it ab initio} DFT with LDA is demonstrated.
$\varsigma$ of Au (111) is evaluated to be -1.86 V, which is in reasonably good
agreement with experimental observations. As compared to other capillary parameters, such as the surface energy, surface stress, and work function, the response is found to vary more strongly with the surface orientation, the value of $\varsigma$ of the (100) surface being only about half of that of the (111) surface.

\acknowledgments
The authors are grateful to Dr. R. Heid and
Dr. K.-P. Bohnen for fruitful discussions and for technical
support during the initial stages of the project.
YU acknowledges support from the Grant-in-Aid for Scientific Research of JSPS (No. 18760082) and
Mechanical Engineering Research Laboratory, Hitachi Ltd.

\end{document}